\documentclass{eptcs}

\usepackage{microtype}
\usepackage{subfig}
\usepackage{amsmath,amsfonts,amsthm}
\usepackage{amssymb}
\usepackage{graphicx}
\usepackage{url} 
  
\usepackage{stmaryrd}
\usepackage{amsmath}
\usepackage{macros} 
\usepackage{breakurl}

\newtheorem*{property}{Property}
\newtheorem{corollary}{Corollary}
\newtheorem{proposition}{Proposition}
\newtheorem{theorem}{Theorem}
\newtheorem{lemma}{Lemma}
\theoremstyle{definition}
\newtheorem{definition}{Definition}
\newcommand{\prgf}[1]{\medskip {} \noindent {\bf #1}}
\title{Labelled $\lambda$-calculi with Explicit Copy and Erase}

\author{Maribel Fern\'{a}ndez, Nikolaos  Siafakas 
\institute{King's College London, Dept. Computer Science, Strand, London WC2R
2LS, UK} 
\email{[maribel.fernandez, nikolaos.siafakas]@kcl.ac.uk}}

\def\titlerunning{Labelled $\lambda$-calculi with explicit copy and erase}
\def\authorrunning{Maribel Fern\'{a}ndez \& Nikolaos  Siafakas}
 
\begin{document}
\maketitle 

\begin{abstract}
  We present two rewriting systems that define labelled explicit
  substitution $\lambda$-calculi. Our work is motivated by the close
  correspondence between L\'evy's labelled $\lambda$-calculus and paths in
  proof-nets, which played an important role in the understanding of the
  Geometry of Interaction.  The structure of the labels in L\'evy's
  labelled $\lambda$-calculus relates to the multiplicative information of
  paths; the novelty of our work is that we design labelled explicit
  substitution calculi that also keep track of exponential information
  present in call-by-value and call-by-name translations of the
  $\lambda$-calculus into linear logic proof-nets.
\end{abstract}

\section{Introduction}
Labelled $\lambda$-calculi have been used for a variety of applications, for
instance, as a technology to keep track of residuals of
redexes~\cite{BarendregtHP:lamcss}, and in the context of optimal reduction,
using L\'evy's labels~\cite{LevyJJ:rceo}. In L\'evy's work, labels give
information about the history of redex creation, which allows the identification
and classification of copied $\beta$-redexes. A different point of view is
proposed in~\cite{AspertiA:pclabels}, where it is shown that a label is actually
encoding a path in the syntax tree of a $\lambda$-term.  This established a tight
correspondence between the labelled $\lambda$-calculus and the Geometry of
Interaction interpretation of cut elimination in linear logic proof-nets
\cite{GirardJY:geoi1i}. 

Inspired by L\'evy's labelled $\lambda$-calculus, we define labelled
$\lambda$-calculi where the labels attached to terms capture reduction traces.
However, in contrast with L\'evy's work, our aim is to use the dynamics of
substitution to include information in the labels about the use of resources,
which corresponds to the exponentials in proof-nets. Exponential structure in proof-nets involves box-structures and connectives that deal with their management, for instance, copying and erasing of boxes. Different translations of the $\lambda$-calculus into proof-nets place boxes at different positions; such choices are also reflected in the paths of the nets.   
In L\'evy's calculus substitution is a meta-operation: substitutions are propagated exhaustively and
in an uncontrolled way. We would like to exploit the fact that substitutions copy
labelled terms and hence paths, but it is difficult to tell with a definition
such as $(MN)^{\alpha}[P/x]=(M[P/x]N[P/x])^{\alpha}$, whether the labels in $P$
are actually copied or not: $P$ may substitute one or several occurrences of a
variable, or it may simply get discarded.

In order to track substitutions we use calculi of explicit substitutions, where
substitution is defined at the same level as $\beta$-reduction.  Over the last
years a whole range of explicit substitution calculi have been proposed, starting
with the work of de Bruijn~\cite{deBruijn:expl} and the
$\lambda\sigma$-calculus~\cite{AbadiM:explsubs}. Since we need to track copy and
erasing of substitutions, we will use a calculus where not only substitutions are
explicit, but also copy and erase operations are part of the syntax.
Specifically, in this paper we use explicit substitution calculi that implement
closed reduction strategies~\cite{FernandezM:clores0,FernandezM:clores}. This may
be thought of as a more powerful form of combinatory
reduction~\cite{HindleyR:cwrlc} in the sense that $\beta$-redexes may be
contracted when the argument part or the function part of the redex is closed.
This essentially allows more reductions to take place under abstractions.  The
different possibilities of placing restrictions on the $\beta$-rule give rise to
different closed reduction strategies, corresponding to different translations of
the $\lambda$-calculus into proof-nets (a survey of available translations can be
found in~\cite{MackieIC:phd}). Closed reduction strategies date back to the late
1980's, in fact, such a strategy was used in the proof of soundness of the
Geometry of Interaction~\cite{GirardJY:geoi1i}.

Labelled $\lambda$-calculi are a useful tool to understand the structure of paths
in the Geometry of Interaction: L\'evy's labels were used to devise optimisations
in GoI abstract machines, defining new strategies of evaluation and techniques
for the analysis of $\lambda$-calculus
programs~\cite{AspertiA:optifp,AspertiA:pill}.  The labels in our calculi of
explicit substitutions contain, in addition to the multiplicative information
contained in L\'evy's labels, also information about the exponential part of
paths in proof-nets.  In other words, our labels relate a static concept---a
path---with a dynamic one: copying and erasing of substitutions. Thus, the labels
can be used not only to identify caller-callee pairs, but also copy and erasing
operations. This is demonstrated in this paper using two different labelled
calculi. In the first system, the $\beta$-rule applies only if the function part
of the redex is closed. We relate this labelled system with proof-nets using the
so-called call-by-value translation. We then define a second labelled
$\lambda$-calculus where the $\beta$-rule applies only if the argument part of
the redex is closed; thus, all the substitutions in this system are closed. We
show that there is a tight relationship between labels in this system and paths
in proof-nets, using the so-called call-by-name translation.

The rest of the paper is organised as follows.  In Section~\ref{sec:background}
we review the syntax of the calculus of explicit substitutions (\lc-terms) and
introduce basic terminology regarding linear logic proof-nets. In
Section~\ref{sec:labelled-terms} we introduce labelled versions of \lc-terms.
Section~\ref{sec:labell-calc-clos} presents the labelled calculus of closed
functions ($\lambda_{lcf}$) that we relate to paths in proof-nets coming from the
call-by-value translation. Similarly, we relate in
Section~\ref{sec:closed-arguments} the labelled version of the calculus of closed
arguments $\lambda_{lca}$ to closed cut-elimination of nets obtained from the
call-by-name translation.  We conclude in Section~\ref{sec:conclusions}.

\section{Background}\label{sec:background}
We assume some basic knowledge of the
$\lambda$-calculus~\cite{BarendregtHP:lamcss}, linear logic~\cite{GirardJY:linl},
and the Geometry of Interaction~\cite{GirardJY:geoi1i}. In this section we recall
the main notions and notations that we will use in the rest of the paper.
\prgf{Labels.} There is a well known connection between labels and paths:
the label associated to the normal form of a term in L\'evy's labelled
$\lambda$-calculus describes a path in the graph of the
term~\cite{AspertiA:pclabels}. The set of labels is generated by the grammar: $
\alpha,\beta:=a \mid \alpha\beta \mid \overline{\alpha} \mid \underline{\alpha}$,
where $a$ is an atomic label. Labelled terms are terms of the $\lambda$-calculus
where each sub-term $T$ has a label attached on it: $T^{\alpha}$. Labelled
$\beta$-reduction is given by $ ((\lambda
x.M)^{\alpha}N)^{\beta}\rightarrow\beta\overline{\alpha}\bullet
M[\underline{\alpha}\bullet N/x]$, where $\bullet$ concatenates labels:
$\beta\bullet T^{\alpha}=T^{\beta\alpha}$. Substitution assumes the variable name
convention \cite{BarendregtHP:lamcss}: 
\[
\begin{array}{lllclll}
x^{\alpha}[N/x] & = & \alpha\bullet N  & \quad & (\lambda y.M)^{\alpha}[N/x] & = & (\lambda y.M[N/x])^{\alpha} \\ 
y^{\alpha}[N/x] & = & y^{\alpha}       & \quad & (MN)^{\alpha}[P/x]          & = & (M[P/x]N[P/x])^{\alpha} \end{array}
\] 
For example, $III$, where $I=\lambda x.x$, can be labelled, and then reduced as follows: 
\\\centerline{\includegraphics[width=0.70\linewidth]{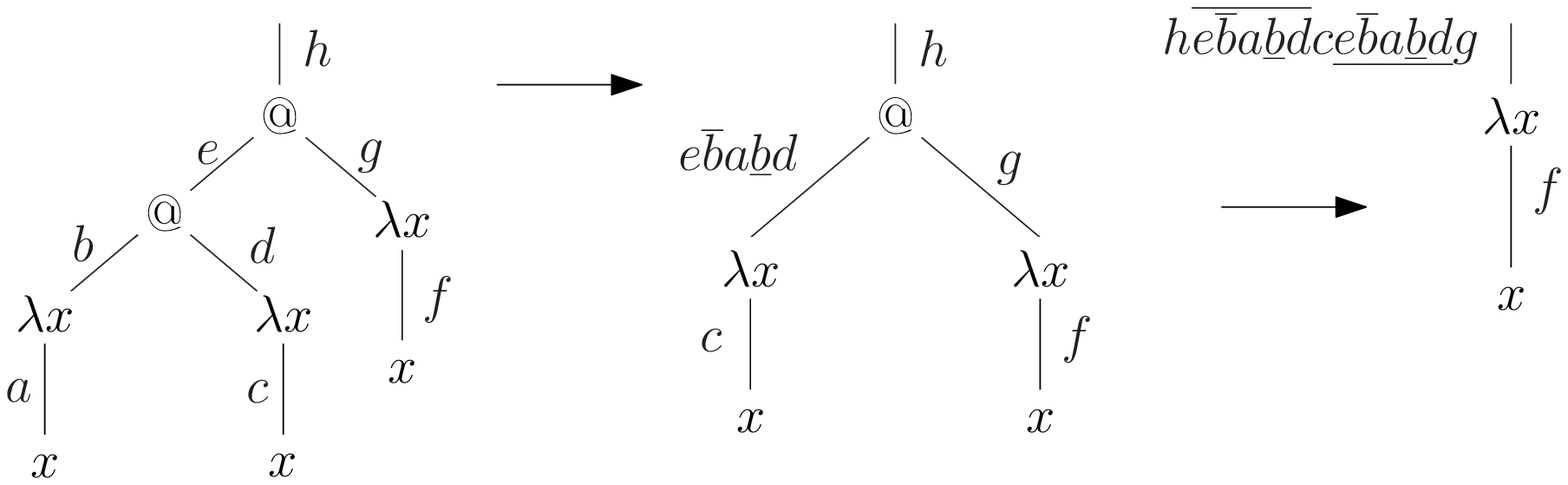} } 
The final label generated describes a path in the tree representation of the 
initial term, if we reverse the underlines. 
Following this path will lead to the sub-term which corresponds
to the normal form, without performing $\beta$-reduction.  This is just one
perspective on Girard's Geometry of Interaction, initially set up to explain
cut-elimination in linear logic. Here we are interested in the
$\lambda$-calculus, but we can use the Geometry of Interaction through a
translation into proof-nets. These paths are precisely the ones that the GoI
Machine follows~\cite{MackieIC:geoim}. In this example, the structure of the
labels (overlining and underlining) tells us about the \emph{multiplicative}
information, and does not directly offer any information about the exponentials.
To add explicitly the exponential information we would need to choose one of the
known translations of the $\lambda$-calculus into proof-nets, and it would be different
in each case.  Further, to maintain this information, we would need to monitor
the progress of substitutions, so we need to define a notion of labelled
$\lambda$-calculus for explicit substitutions. The main contribution of this
paper is to show how this can be done.

\prgf{Explicit substitutions and resource management.}
Explicit substitution calculi give first class citizenship to the otherwise
meta-level substitution operation.  Since we need to track copy and erasing of
substitutions, in this paper we will use a calculus where not only substitutions
are explicit, but also copy and erase operations are part of the syntax.  The
explicit substitution calculi defined
in~\cite{FernandezM:clores0,FernandezM:clores} are well-adapted for this work:
besides having explicit constructs for substitutions, terms include constructs
for copying ($\delta$) and erasing ($\epsilon$) of substitutions.  The motivation
of such constructs can be traced back to linear logic, where the structural rules
of weakening and contraction become first class logical rules, and to Abramsky's
work~\cite{AbramskyS:comil} on proof expressions. 

The table below defines the syntax of \lc-terms, together with the variable
constraints that ensure that variables occur linearly in a term. We use
$\FV(\cdot)$ to denote the set of free variables of a term.  We refer the reader
to~\cite{FernandezM:clores} for a compilation from $\lambda$-terms to \lc-terms.
\begin{center}
$\begin{array}{lll}
  \LINEDECOA                                                                                                 \\
  \mbox{Term}        & \mbox{Variable Constraint}                         & \mbox{Free variables}            \\
  \LINEDECOA                                                                                                 \\
  x                  & \VOID                                              & \SET{x}                          \\
  \lambda x.M        & x\in\FV(M)                                         & \FV(M) - \SET{x}                 \\
  MN                 & \FV(M)\cap\FV(N)=\emptyset                         & \FV(M)\cup\FV(N)                 \\
  \epsilon_{x}.M     & x\not\in\FV(M)                                     & \FV(M)\cup \SET{x}               \\
  \delta_{x}^{y, z}.M & x\not\in \FV(M), y\not=z, \SET{y,z}\subseteq\FV(M) &
  (\FV(M)- \SET{y,z}) \cup \SET{x} \\ M[N/x]             & x\in\FV(M), (\FV(M)-\SET{x})\cap \FV(N)=\emptyset  & (\FV(M)-\{ x\})\cup\FV(N)        \\
  \LINEDECOB
  \end{array}$
\end{center}

For example, the compilation of $\lambda x.\lambda y.x$ is $\lambda x.\lambda y.\epsilon_{y}.x$, 
and the compilation of $(\lambda x.xx)(\lambda x.xz)$ is $(\lambda x.\delta_{x}^{x', x''}.x'x'')(\lambda x.xz)$.
We remark that both yield terms that satisfy the variable constraints.

Using \lc-terms, two explicit substitution calculi were defined
in~\cite{FernandezM:clores}: $\lambda_{cf}$, the calculus of closed functions,
and $\lambda_{ca}$, the calculus of closed arguments. In the former, the
 $\beta$-rule requires the function part of the redex to be closed, whereas in the latter,
the argument part must be closed to trigger a $\beta$-reduction. In the
rest of the paper we define labelled versions of these calculi and relate labels
to proof-net paths.

\begin{figure}
\centerline{
  \subfloat[A proof-net obtained by the call-by-value translation of the 
   term $(\lambda xy.xy)(\lambda x.x)$]
    { 
     \label{fig:exanetA}\qquad\includegraphics[width=0.25\linewidth]{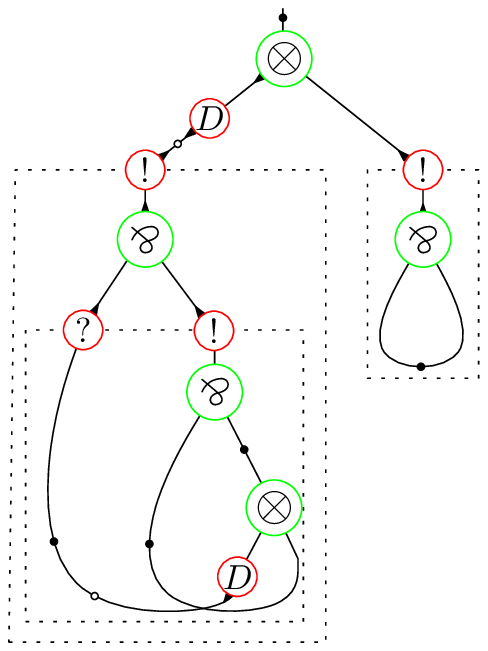}\qquad 
    } \qquad \qquad 
    \subfloat[A bouncing (1), a twisting (2), and a straight path (3)]
    {
      \label{fig:exanetB}
      \qquad\includegraphics[width=0.36\linewidth]{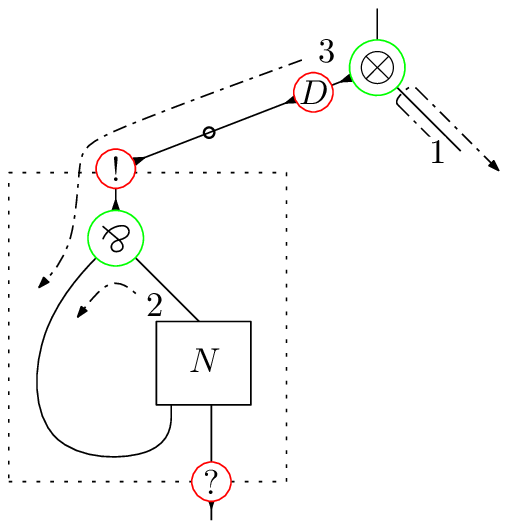}\qquad
    }
}\vspace{-10pt}
\caption{Examples of proof-nets and paths}
\label{fig:exanet}
\vspace{-15pt}
\end{figure}

\prgf{Proof-nets and the Geometry of Interaction.}
The canonical syntax of a linear logic proof is a graphical one: a proof-net.
Nets are built using the following set of nodes: \emph{Axiom ($\bullet$)} and
\emph{cut ($\circ$)} nodes (such nodes are dummy-nodes in our work and simply
represent ``linking'' information); Multiplicative nodes: $\otimes$ and $\lpar$;
Exponential nodes: \emph{contraction (fan)}, \emph{of-course (!)}, \emph{why-not
(?)}, \emph{dereliction (D)} and \emph{weakening (W)}.  A sub-net may be enclosed
in a \emph{box} built with one \emph{of-course}-node and $n\ge 0$
\emph{why-not}-nodes, which we call auxiliary doors of a box.
The original presentation of a net is oriented so that edges designated as
conclusions of a node point downwardly: we abuse the natural orientation and
indicate with an arrow-head the conclusion of the node (see
Figure~\ref{fig:exanet}); all other edges are premises.
Such structures may be obtained by one of the standard translations, call-by-name
or call-by-value \cite{MackieIC:phd,GirardJY:linl}, of $\lambda$-terms.  For
instance, the net in Figure~\ref{fig:exanetA} is obtained by the call-by-value
translation of the $\lambda$-term $(\lambda xy.xy)(\lambda x.x)$.

We assume a weak form of cut elimination---closed cut elimination--- denoted by
$\Rightarrow$. This is ordinary multiplicative and exponential proof-net
reduction with the restriction that every exponential step can handle only boxes
with no auxiliary doors. We refer the reader
to~\cite{GirardJY:geoi1i,MackieIC:phd} for a specification of this
cut-elimination strategy, which is also used in the context of Interactions
Nets~\cite{MackieIC:IntNets}.

In this work we obtain labelled (weighted) versions of nets via inductive
translations of \lc-terms, which we define later on.

\prgf{Weighted nets.} To each edge of a net, we associate a \emph{weight}
($w$) built from terms of the dynamic algebra $L^\star$: constants $p,q$
(multiplicative), $r,s,t,d$ (exponential), $0$ and $1$; an associative
composition operator $``."$ with unit $1$ and absorbing element $0$; an involution
$(\cdot)^{*}$ and a unary morphism $!(\cdot)$. We use meta-variables $\alpha,
\beta, \dots$ for terms and we shall write $!^n(\cdot)$ for $n \ge 0$
applications of the morphism.  Intuitively, weights are used to identify paths,
and the algebra is used to pick out the paths that survive reduction.  We omit
the definition of the correct labelling (with weights) of a net: this will again
be obtained by translations of $\lambda_c$-terms into labelled proof-nets.

We use metavariables $\phi, \chi \dots$ to range over paths. Paths are assumed to
be {\bf a)} \emph{non-twisting}, that is, paths are not over different premises
of the same node and {\bf b)} \emph{non-bouncing}, that is, paths do not bounce
off nodes. We call such paths \emph{straight}; these traverse a weighted edge $e$
forwardly when moving towards the premise of the incident node (resp. backwardly
$e^r$ when moving towards the conclusion) such that direction changes happen only
at cut and axiom links. For instance, Figure \ref{fig:exanetB} shows a bouncing,
a twisting, and a straight path respectively.
The weight of a path is $1$ if it traverses no edge --- this weight is the
identity for composition which we usually omit; if $\phi=e\cdot\psi$ is a path
then its weight $w(\phi)$ is defined to be\footnote{ Traditionally, weights are
composed antimorphically but
  in this work we read paths and compose weights from left to right.} $
w(e) \cdot w(\psi) $ and we have $w(e^r)=w(e)^*$. We are mainly interested in the
statics of the algebra and omit the equations that terms satisfy.  We refer the
reader to~\cite{Danos:phd,AspertiA:pill,AspertiA:optifp} for a more detailed
treatment.

\section{Labelled terms}\label{sec:labelled-terms}
We attach labels to \lc-terms in order to capture information not only about
$\beta$-reductions but also about propagation of substitutions. We adopt the same
language for labels as in~\cite{Siafakas:Full}, where a confluent system
($\lambda_{lcf}$) is defined and informally related to traces in a call-by-value
proof-net translation.  To establish the required correspondence, in this paper
we translate (in Sections~\ref{sec:labell-calc-clos} and
\ref{sec:closed-arguments}) our labels into terms of $L^\star$ (defined in the previous section), which is the
de-facto language for labels in proof-nets.

\begin{definition}
  Labels are defined by the following grammar; $a$ is an atomic label taken from
  a denumerable set \SET{a,b, \dots}, and all labels in $C$ are atomic. 
 $$\alpha,\beta:=a \mid \alpha \cdot \beta \mid \overline{\alpha} \mid
  \underline{\alpha} \mid C;\qquad C:=\overrightarrow{E} \mid \overleftarrow{E};
  \qquad E:=D \mid {!} \mid ?  \mid R \mid S \mid W;$$
\end{definition}
These labels are similar to L\'{e}vy's labels except that we have (atomic)
markers to describe exponentials, motivated by the constants in the algebra
$L^\star$. Atomic labels from \SET{a,b,\dots} correspond to $1$ while $W$
corresponds to $0$; the marker $?$ corresponds to $t$ and \SET{d,r,s} are easily
recognised in our labels. Multiplicative constants \SET{p,q} are treated
implicitly via over-lining and underlining. One may recover the multiplicative
information, which is simply a bracketing,  using the function $f$ which is the
identity transformation on all labels except for $f ~\overline{\alpha}=
\overrightarrow{Q} \cdot (f ~ \alpha) ~ \cdot
\overleftarrow{Q}$ and $f ~\underline{\alpha}= \overrightarrow{P}
\cdot (f ~ \alpha) ~ \cdot \overleftarrow{P}$. The marker $!$ deserves more
attention since the straightforward analogue in the algebra is the morphism
$!(\cdot)$. The purpose of the marker is to delimit regions in the label, that
is, paths that traverse edges entirely contained in a box.

\prgf{Initialisation:}
To each term in \lc, except for $\delta$, $\epsilon$ and substitution-terms, we
associate a unique and pairwise distinct label from \SET{a,b, \cdots}. Notice
that we do not place any exponential markers on initialised terms: it is the
action of the substitution that yields their correct placement.  From now on, we
assume that all \lc-terms come from the compilation of a $\lambda$-term and
receive an initial labelling. Free and bound variables of labelled
$\lambda_c$-terms are defined in the usual way.

\section{The labelled calculus of closed functions}\label{sec:labell-calc-clos} 
In this section we define the labelled calculus $\lambda_{lcf}$, that yields
traces for the call-by-value translation of \lc-terms into linear logic proof
nets.

\begin{definition} [Labelled Reduction in $\lambda_{lcf}$]
\label{def:lab-red-lcf}
The \emph{Beta}-rule of the labelled calculus $\lambda_{lcf}$ is
defined by
$$((\lambda
  x.M)^{\alpha}N)^{\beta} \cfmer \beta \overstufflcf \bullet
  M[\understufflcf \bullet N/x] ~~~~~ \textit{if} ~\FV((\lambda x.M)^{\alpha})=\emptyset$$
The operator $\bullet$ and the function $(\cdot)^{r}$ on labels are defined in Table~\ref{tab:bulletandrev}.
\begin{table}\vspace{-5pt}
\[
\begin{array}{llllllll}
  (a)^{r}                  & = & a                             & \qquad &                  & \beta\bullet x^{\alpha}            & = & x^{\beta\alpha} \\ 
  (\alpha\beta)^{r}        & = & (\beta)^{r}\cdot(\alpha)^{r}  &        &                  & \beta\bullet(\lambda x.M)^{\alpha} & = & (\lambda x.M)^{\beta\alpha}\\ 
  (\overline{\alpha})^{r}  & = & \overline{(\alpha)^{r}}       &        &                  & \beta\bullet(MN)^{\alpha}          & = & (MN)^{\beta\alpha}\\ 
  (\underline{\alpha})^{r} & = & \underline{(\alpha)^{r}}      &        &                  & \alpha\bullet(\delta_{x}^{y, z}.M) & = & (\delta_{x}^{y, z}.\alpha\bullet M)\\ 
  (\overrightarrow{E})^{r} & = & \overleftarrow{E}             &        &                  & \alpha\bullet(\epsilon_{x}.M)      & = & (\epsilon_{x}.\alpha\bullet M)\\ 
  (\overleftarrow{E})^{r}  & = & \overrightarrow{E}            &        &                  & \alpha\bullet (M[N/x])             & = & (\alpha \bullet M)[N/x] 
\end{array}
\]\vspace{-15pt}
\caption{\label{tab:bulletandrev} Operation of $(\cdot)^r$ and $\cdot\bullet\cdot$}
\end{table}
We place substitution rules ($\sigma$) at the same level as the \emph{Beta}-rule.
These are given in Table~\ref{tab:lcfsigma}.
\begin{table}\vspace{-0pt}
  \def \LINEDECOA {\hline \hspace{0.15\linewidth} & &
    \hspace{0.10\linewidth} &  \hspace{0.35\linewidth}  \vspace{-9pt}}
  \def \LINEDECOB {\vspace{-9pt} \\\hline}
  \center \begin{tabular}{llcll}
    \LINEDECOA \\
    \multicolumn{1}{l}{Rule}&
    \multicolumn{3}{c}{Reduction~~~~~~~~~~~~~~~~~~~~}& 
    Condition\\
    \LINEDECOA\\
    \emph{Lam}&
    $(\lambda y.M)^{\alpha}[N/x]$&
    $\cfmer$&
    $(\lambda y.M[\wright\bullet N/x])^{\alpha}$&
    $\FV(N)=\emptyset$\\
    \emph{App1}&
    $(MN)^{\alpha}[P/x]$&
    $\cfmer$&
    $(M[P/x]N)^{\alpha}$&
    $x\in\FV(M)$\\
    \emph{App2}&
    $(MN)^{\alpha}[P/x]$&
    $\cfmer$&
    $(MN[P/x])^{\alpha}$&
    $x\in\FV(N)$\\
    \emph{Cpy1}&
    $(\delta_{x}^{y, z}.M)[N/x]$&
    $\cfmer$&
    $M[\rright\bullet N/y][\sright\bullet N/z]$&
    $\FV(N)=\emptyset$\\
    \emph{Cpy2}&
    $(\delta_{x}^{y, z}.M)[N/x']$&
    $\cfmer$&
    $(\delta_{x}^{y, z}.M[N/x'])$&
    \VOID\\
    \emph{Ers1}&
    $(\epsilon_{x}.M)[N/x]$&
    $\cfmer$&
    $M,\qquad\{\overrightarrow{\sf W}\bullet N\}\cup B$&
    $\FV(N)=\emptyset$\\
    \emph{Ers2}&
    $(\epsilon_{x}.M)[N/x']$&
    $\cfmer$&
    $(\epsilon_{x}.M[N/x'])$&
    \VOID\\
    \emph{Var}&
    $x^{\alpha}[N/x]$&
    $\cfmer$&
    $\alpha\bullet N$&
    \VOID\\
    \emph{Cmp}&
    $M[P/y][N/x]$&
    $\cfmer$&
    $M[P[N/x]/y]$&
    $x\in\FV(P)$\\
    \LINEDECOB
\end{tabular}
\caption{\label{tab:lcfsigma}Labelled substitution ($\sigma$) rules in $\lambda_{lcf}$}
\vspace{-15pt}
\end{table}
We write $\to_{lcf}^{*}$ for the transitive reflexive closure of $\to_{lcf}$ and
we may omit the name of the relation when it is clear from the context. Reduction
is allowed to take place under any context satisfying the conditions.
\end{definition}

The calculus defined above is a labelled version of $\lambda_{cf}$, the calculus
of closed functions~\cite{FernandezM:clores}. The conditions on the rules may not
allow a substitution to fully propagate (i.e., a normal form may contain
substitutions) but the calculus is adequate for evaluation to weak head normal
form~\cite{FernandezM:clores}. Although reduction is weak, the system does not
restrict reduction under abstraction altogether as do theories of the weak
$\lambda$-calculus. 

Intuitively, the purpose of the \emph{Beta}-rule is to capture two paths, one
leading into the body and one to the argument of the proof-net representation of
the function application. The controlled copying and erasing (\emph{Cpy1} and
\emph{Ers1}) of substitutions allows the identification of paths that start from
contraction nodes and weakening nodes respectively.  Rule \emph{Ers1} has a side
effect; erased paths are kept in a set $B$, which is initially empty. Formally,
this rewriting system is working on pairs $(M,B)$ of a $\lambda$-term and a set
of labels. The set $B$ deserves more regards: there exist paths in the GoI that
do not survive the action of reduction (they are killed off by the dynamics of
the algebra), however, one could impose a strategy where such sub-paths may be
traversed indeed.  For instance, evaluating arguments before function application
gives rise to traversals of paths that lead to terms that belong in $B$. In this
sense, it is not only the arguments that get discarded but also their labels,
i.e. the paths starting from variables that lead to unneeded arguments. We omit
explicit labels on copying ($\delta$) and erasing ($\epsilon$) constructs: these
are used just to guide the substitutions.  The composition rule \emph{Cmp} is vital in the
calculus because we may create open substitutions in the \emph{Beta}-rule. 

The calculus is $\alpha$-conversion free: the propagation of open substitutions
through abstractions is the source of variable capture, which is here avoided due
to the conditions imposed on the \emph{Lam}-rule. Moreover, the labelled calculus
has the following useful properties:
\begin{property}
\begin{enumerate}
\item \emph{Strong normalisation of substitution rules}. The reduction
  relation generated by the $\sigma$-rules is terminating.
\item \emph{Propagation of substitutions}.  Let $T=M[N/x]$. If $\FV(N)
  = \emptyset$ then $T $ is not a normal form. As a corollary closed
  substitutions can be fully propagated.
\item \emph{Confluence}. $\lambda_{lcf}$ reductions are confluent: if
  $M \cfmer^* N_1$ and $M \cfmer^* N_2$ then there exists a term $P$
  such that $N_1 \cfmer^* P$ and $N_2 \cfmer^* P$.
\end{enumerate}
\end{property}
The proof of termination of $\sigma$ is based on the observation that rules push
the substitution down the term (which can be formalised using the standard
interpretation method).  Since the propagation rules are defined for closed
substitutions, it is easy to see that closed substitutions do not block. The
proof of confluence is more delicate. We derive confluence in three steps: First,
we show confluence of the $\sigma$-rules (local confluence suffices, by Newman's
lemma~\cite{NewmanMHA:thecd}, since the rules are terminating). Then we show that
$\beta$ alone is confluent, and finally we use Rosen's
lemma~\cite{RosenBK:trems}, showing the commutation of the $\beta$ and $\sigma$
reduction relations. For detailed proofs we refer the reader
to~\cite{SiafakasN:PhD}.

\subsection*{\texorpdfstring{Labels in $\lambda_{lcf}$ and paths in the call-by-value translation}
                           {Labels in  lambda-lcf     and paths in the call-by-value translation}
           }
There is a correspondence between our labels and the paths in weighted proof
nets. The aim of the remainder of the section is to justify the way in which this
calculus records paths in proof-nets.  This will provide us with a closer look at
the operational behaviour of the calculus, especially at the level of propagation
of substitutions, and will highlight the relationship, but also the differences,
between term reduction and proof-net reduction.

We first define a call-by-value translation from labelled $\lambda_{c}$-terms to
proof-nets labelled with weights taken from algebra $L^*$, and then we show that
the set of labels generated at each rewrite step in the calculus coincides with
the set of weights in the corresponding nets.  For simplicity, we first consider
the translation of unlabelled terms; then we extend the translation to labelled
terms.

\begin{definition}\label{def:transcbv}
  In Figure~\ref{cap:Translationofcterms} we give the translation function
  $G_n(\cdot)$ from unlabelled $\lambda_{c}$-terms to call-by-value proof-nets.
  We use a parameter $n$ ($n\ge 0$) to record a current-box level, which
  indicates the box-nesting in which the translation works. The translation of a
  term $M$ is obtained with $G_0(M)$, indicating the absence of box-nesting in
  initial terms.  In our translation we omit the weights $1$ of cuts and axioms.
  In general, the translation of a term $M$ at level $n$ is a proof-net:
  \\\centerline{\includegraphics[width=0.1\textwidth]{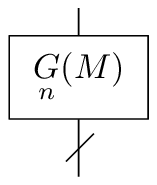}} where the crossed
  wire at the bottom represents a set of edges corresponding to the free
  variables in $M$.

  \begin{figure}[t]\vspace{-10pt}\centering 
    \includegraphics[%
    width=0.9\textwidth]{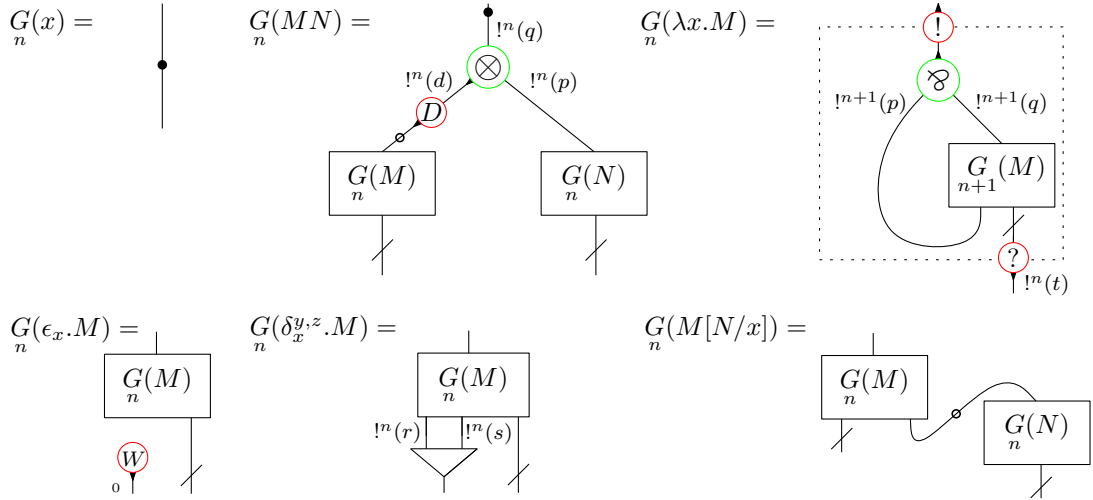} 
    \caption{\label{cap:Translationofcterms}Translation of 
      $\lambda_{c}$-terms into weighted call-by-value proof-nets}
    \vspace{-15pt}
  \end{figure}
\end{definition}
 
At the syntactical level, the correspondence of a $\delta$-term to a duplicator
(fan) node is evident as well as the correspondence of $\epsilon$-term to
weakening $(W)$ nodes. In the former case, we assume that the variable $y$ (resp.
$z$) corresponds to the link labelled with $r$ (resp. $s$). The free variable $x$
corresponds to the wire at the conclusion of the fan. The case for the erasing
term is similar, the free variable $x$ corresponds to the conclusion of the
weakening node. The encoding of a substitution term is the most interesting: a
substitution redex corresponds to a cut in proof-nets.

We next give the translation of labelled terms, which is similar except that now
we must translate labels of the calculus into terms of the algebra. This is a
two-step process: first we must consider how labels correspond to weights in
proof-nets, and then we must place the weight on an edge of the graph.

\begin{definition} \label{def:translabel}
  The weight of a label is obtained by the function $lw:\!:label \to level
  \to(weight,level)$ where $level \in \mathbb{N}$ is a level (or box depth)
  number. Given a label and the level number of the first label, the function
  defined in Table~\ref{tab:toliw} yields a weight together with the level number
  of the last label. We assume that the input level number always is an
  appropriate one.

  \begin{table}
  \[
  \begin{array}{cc|cc}
    \mbox{Atomic labels} & &  \mbox{Composite labels}\\ \hline &&&\\
    \begin{array}{lcl}
      \elstar[a] n                        & = & (1,n)              \\
      \elstar[\overrightarrow{\sf{R}}]  n & = & (!^n(r),n)         \\
      \elstar[\overleftarrow{\sf{R}}]  n  & = & (!^n(r^*),n)       \\
      \elstar[\overrightarrow{\sf{S}}]  n & = & (!^n(s),n)         \\
      \elstar[\overleftarrow{\sf{S}}]  n  & = & (!^n(s^*),n)       \\
      \elstar[\overrightarrow{\sf{D}}]  n & = & (!^n(d),n)         \\
      \elstar[\overleftarrow{\sf{D}}]  n  & = & (!^n(d^*),n)       \\
      \elstar[\overrightarrow{\sf{?}}]  n & = & (!^{n-1}(t^*),n-1) \\
      \elstar[\overleftarrow{\sf{?}}]  n  & = & (!^{n}(t) ,n+1)    \\
      \elstar[\overrightarrow{\sf{!}}]  n & = & (1,n-1)            \\
      \elstar[\overleftarrow{\sf{!}}]  n  & = & (1,n+1)
    \end{array} & &
    \begin{array}{lll}
      \elstar[\overline{\alpha}] n &=& ((!^n(q)) \cdot w \cdot !^{n'}(q^*) ,n') \\
      &\mbox{where}& (w ,n') = \elstar[\alpha] n  \\
      \\
      \elstar[\underline{\alpha}] n &=& ((!^n(p)) \cdot w \cdot !^{n'}(p^*) ,n') \\
      &\mbox{where}& (w ,n') = \elstar[\alpha] n  \\
      \\
      \elstar[\alpha \beta] n &=& (w \cdot w' ,n'')  \\
      &\mbox{where}& (w  , n') = \elstar [\alpha] n  \\
      &            & (w' , n'') = \elstar[\beta] {n'}
    \end{array}
  \end{array}
  \]\vspace{-10pt}
  \caption{Translation of labels into weights}
  \label{tab:toliw}
\vspace{-10pt}
\end{table}
\end{definition}

Before we give the translation, we introduce a convention that will help us to
reason about input and output levels: instead of projecting the weight from the
tuple in the previous definition, we place the tuple itself on a wire like this:
\\\centerline{\includegraphics[width=0.35\textwidth]{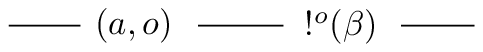}} where $\alpha$ is the weight of the
translated label, $o$ is a level number. Now, after projecting the weight from
the tuple one may compose with existing weights on the wire ($!^o(\beta)$). This
simply introduces a delay in our construction that  helps us to maintain the
levels.

\begin{figure}
\centerline{\includegraphics[width=0.8\textwidth]{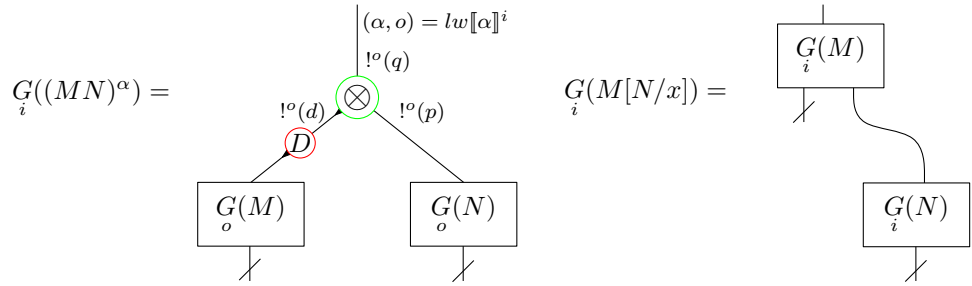}}
\vspace{-5pt}
\caption{Translation of labelled application and substitution terms}
\label{fig:translabgraph} 
\vspace{-10pt}
\end{figure}

The translation of a labelled term is obtained using the function $G_i(M)$, which
is the same as in Definition \ref{def:transcbv}, with the difference that now
when we call $G_i(M)$, the parameter $i$ depends on the translation of the label.
Specifically, for each term that has a label on its root, we first translate the
label using Definition ~\ref{def:translabel}, and place the obtained output onto the 
root of $G$. In Figure~\ref{fig:translabgraph} we show the translation of labelled application and substitution terms
and the remaining cases can be easily reconstructed from the translation of the
unlabelled terms.
Thus, the only difference is that when a term has a label on the root, the translation
must use the output level to propagate to subterms.

To extract the external label of a term we use the function:
\[
\begin{array}{lclclcl}
  \LABEL x^\alpha        & = & \alpha   & \quad & \LABEL~(MN)^\alpha        & = & \alpha   \\
  \LABEL (\lambda x.M)^a & = & \alpha   & \quad & \LABEL (\delta_x^{y,z}.M) & = & \LABEL M \\
  \LABEL (\epsilon_x.M)  & = & \LABEL M & \quad & \LABEL (M[N/x])           & = & \LABEL M
\end{array}
\]
Similarly, we can get the label of a free variable in a term. This is just a
search and we omit the definition. Thanks to the linearity of terms, there is
exactly one wire for each free variable in the term.

\prgf{Main result.}
Before proving the main result of this section (Theorem~\ref{th:main1}), which
states the correspondence between labels and paths, we need a few general
properties.
\begin{proposition}[First and last atomic
  labels]\label{lem:firstlabel}\label{lem:lastlabel}
  Let $k$ be the external label of an initialised term $T$, and assume
  $T \to^* T'$.
  \begin{enumerate}
  \item If $(label~ T') = l_1 \dots l_n$, $n \ge 1$, then $l_1$ is $k$.
  \item Let $N$ be an application, abstraction or variable subterm of
    $T'$ with $(label\ N) = l_1 \dots l_n$, $n \ge 1$. The atomic label $l_n$
    identifies an application (resp. abstraction, resp. variable) term in $T$ iff
    $N$ is an application (resp. abstraction, resp. variable) term.

  \end{enumerate}
\end{proposition}
This property captures the idea that we cannot lose the original root of a
reduction and terms never forget about their initial label. Notice that we
consider terms that do receive a label by the initialisation. Thus the last
atomic label of a string on a term-construct is the label the construct has
obtained by initialisation. This is because labels get prefixed by the actions of
the calculus. Additionally, notice that in $((\lambda x. M)^\alpha N)^\beta \to \beta
\overline{\alpha} \bullet M[ \underline{\alpha^r} N/z]$, where we forget about
the markers, we know that the last label of $\beta$ must be the one of the
application node in which $M$ was the functional part. But we also know the first
label of $\alpha$: if it is atomic, then it is the label of this $\lambda$ in the
initial term. Otherwise it is the label of the functional edge of the application
node identified by the last label of $\beta$.  One argues similarly for the
argument.
\begin{lemma}\label{lem:brights}
  If $T = M[N/x]$ then there is a decomposition $(label\ N) = \omega
  \underline{\alpha}\sigma $ such that $\omega$ is a prefix built with
  exponential markers having the shape $\overrightarrow{E}_1 \dots
  \overrightarrow{E}_n $ with $n \ge 0$.
\end{lemma}
 \begin{proof}
   A simple inspection of the rewrite rules shows that we always prefix the
   external label of $N$ with some exponential marker as long as the substitution
   propagates with label sensitive rules. The moment where $\omega$ itself gets
   prefixed is during a variable substitution which stops the process.
 \end{proof}
The previous statement allows us to point out the distinguishing
pattern of labels on substitution terms where we shall see that label sensitive
propagation of substitutions corresponds to building an exponential path in a
proof-net.

\begin{corollary}
  The atomic label that stands on an initialised variable can be
  followed only by $\omega \underline{\alpha}\sigma$.
\end{corollary}
 \begin{proof}
   This is a consequence of the last-label property stated above and the action
   of the rewrite rule \emph{Var}.
 \end{proof}

 Next we show that the labels of the calculus adequately trace paths in
 proof-nets.  In particular, we show that the set of labels generated in the
 calculus coincides with weights of straight paths in proof-nets in the following
 sense:
 
\begin{theorem}\label{th:main1}
  Let $W_G=\{w(\phi) \mid \phi \in \mbox{straight paths of } G\}$
  denote the set of weights of straight paths observable in a graph $G$ and let $T$ be a
  labelled term. If $T \to T'$ then $W_{G(T)} = W_{G(T')}$.
\end{theorem}
\begin{proof}
  The proof is by induction on $T$ and is given in the appendix.
\end{proof}
\section{The labelled calculus of closed arguments}\label{sec:closed-arguments}
In this section we define a labelled calculus, called $\lambda_{lca}$,
that yields traces for the call-by-name translation of \lc-terms into
linear logic proof-nets.
\def \camer {\to_{lca}}
\def \overstufflca{  \overline{\alpha}   }
\def \understufflca{  \underline{(\alpha)^{r}}\bleft   }
\def \thebetalca{((\lambda x.M)^{\alpha}N)^{\beta} \camer \beta \bullet \overstufflca
M[ \understufflca \bullet N/x] ~~~~~\textit{if}~ \FV(N)=\emptyset}
\begin{definition}[Labelled reduction in $\lambda_{lca}$]
The new \emph{Beta}-rule is defined by:
\\\centerline{$\thebetalca$}
\noindent where the operator $\bullet$ is given in
Definition~\ref{def:lab-red-lcf}. The substitution rules for this system are
presented below:
\begin{center}
  \def \LINEDECOA {\hline \hspace{0.17\linewidth} & &
    \hspace{0.10\linewidth} &  \hspace{0.35\linewidth}  \vspace{-9pt}}
  \def \LINEDECOB {\vspace{-5pt} \\\hline}
  \begin{tabular}{llcll}
    \LINEDECOA \\
    \multicolumn{1}{l}{Rule}&
    \multicolumn{3}{c}{Reduction~~~~~~~~~~~~~~~~~~~~}& 
    Condition\\
    \LINEDECOA\\
    \emph{Lam}&
    $(\lambda y.M)^{\alpha}[N/x]$&
    $\rightarrow_{\lambda_{lca}}$&
    $(\lambda y.M[N/x])^{\alpha}$&
    \VOID\\
    \emph{App1}&
    $(MN)^{\alpha}[P/x]$&
    $\camer$&
    $(M[P/x]N)^{\alpha}$&
    $x\in\FV(M)$\\
    \emph{App2}&
    $(MN)^{\alpha}[P/x]$&
    $\camer$&
    $(MN[\wright  \bullet P/x])^{\alpha}$&
    $x\in \FV(N)$\\
    \emph{Cpy1}&
    $(\delta_{x}^{y, z}.M)[N/x]$&
    $\camer$&
    $M[\rright  \bullet N/x][\sright \bullet N/x]$&
    \VOID\\
    \emph{Cpy2}&
    $(\delta_{x}^{y, z}.M)[N/x']$&
    $\camer$&
    $(\delta_{x}^{y, z}.M[N/x'])$&
    \VOID\\
    \emph{Ers1}&
    $(\epsilon_{x}.M)[N/x]$&
    $\camer$&
    $M,\qquad\{\overrightarrow{\sf W}\bullet N\}\cup B$&
    \VOID\\
    \emph{Ers2}&
    $(\epsilon_{x}.M)[N/x']$&
    $\camer$&
    $(\epsilon_{x}.M[N/x'])$&
    \VOID\\
    \emph{Var}&
    $x^{\alpha}[N/x]$&
    $\camer$&
    $\alpha\dright \bullet N$&
    \VOID\\
    \LINEDECOB
  \end{tabular}
\end{center}
\end{definition}
This is the labelled version of the calculus of closed arguments in
\cite{FernandezM:clores} and we refer the reader to \cite{SiafakasN:PhD} for a
proof of confluence for the labelled version.

\subsection*{\texorpdfstring{Labels in $\lambda_{lca}$ and paths in the call-by-name translation}
                           {Labels in   lambda-lca    and paths in the call-by-name translation} 
           } 
The particularities of the calculus are best understood via the correspondence to
the call-by-name translation. Thus, let us move directly to the translation of
terms into call-by-name proof-nets.  We provide a simplified presentation where
instead of placing a translated label on a wire, we simply place the label
itself.  As before, multiplicative information is kept implicit via overlining
and underlining. Hence, the general form of the translation takes a labelled term
and places the label of the term (when it has one) at the root of the graph:
\begin{figure}\vspace{-5pt}
  \centering
  \includegraphics[width=1\linewidth]{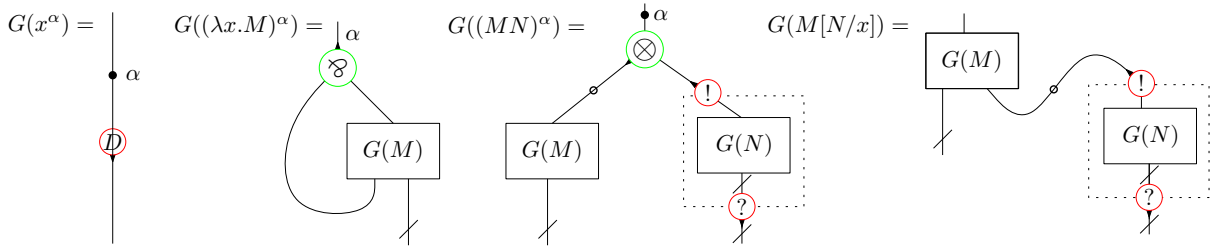} 
  \caption{Translation of $\lambda_{c}$ into call-by-name nets} 
  \label{fig:lcbv}\vspace{-15pt}
\end{figure}
We give the translation in Figure~\ref{fig:lcbv}. Remark that the translation of
an argument (and the substitution) always involve a box structure. We do not
repeat translations for $\delta$ and $\epsilon$-terms since these translate in
the same way as before. 

The following theorem is the main result of this section. It establishes a
correspondence between labels and paths in the call-by-name proof-net
translation.
\begin{theorem} \label{thm:cbn}
\begin{enumerate}
\item If $T \to_{lca} T'$ then $W_{G(T)} = W_{G (T')}$, where $W_G$ is
  defined as in Theorem~\ref{th:main1}.
\item Suppose $T$ is a term obtained by erasing the labels and
  $\to_{ca}$ is the system generated by the rules for $\lambda_{lca}$
  by removing the labels.  If $T \to_{ca} T'$ then $G(T) \Rightarrow^*
  G(T')$ using closed cut elimination, where $G$ is the
  call-by-name translation.
\end{enumerate}
\end{theorem}
This theorem is stronger than Theorem~\ref{th:main1}: there is a correspondence
between labels in $\lambda_{lca}$ and paths in the call-by-name proof-net
translation, and between the dynamics of the calculus and the proof-net dynamics
of closed cut elimination $\Rightarrow$ (due to space constraints we omit the
definition of $\Rightarrow$ and refer to~\cite{SiafakasN:PhD}).  Rules
\emph{App1,
  Lam, Cpy2 and Ers2} correspond to identities while the remaining
rules correspond to single step graph rewriting. To obtain a similar result for
$\lambda_{lcf}$ with the call-by-value translation, we need to impose more
conditions on the substitution rules (requiring closed values instead of simply
closed terms).

\section{Conclusions and future work}\label{sec:conclusions}
We have investigated labelled $\lambda$-calculi with explicit substitutions.  The labels give insight into how the dynamics of a
$\lambda$-calculus corresponds to building paths in proof-nets, and  they
also allow us to understand better the underlying calculi. For instance, label
insensitive substitution rules witness that some actions in the calculi capture
``less essential'' computations, that is, an additional price for bureaucracy of
syntax is paid in relation to the corresponding proof-net dynamics.  From this
point of view, it is interesting to ask about whether strategies for these
calculi exist such that each propagation of substitution corresponds to
stretching a path in a corresponding proof-net. On the other hand, investigation
of new labelled versions of known strategies defined for the underlying calculi
could help in understanding and establishing requirements for new proof-net
reduction strategies.

The use of closed reduction in the current work simplifies the computation of the
labels, since only closed substitutions are copied/erased. The methodology can be
extended to systems that copy terms with free variables, but one would need to
use global functions to update the labels; instead, using closed reduction, label
computations are local, in the spirit of the Geometry of Interaction. Notice that
duplication (resp. erasing) of free variables causes further copying (resp.
erasing), which in our case would require on the fly instantiation of additional
$\delta$-terms (resp. $\epsilon$-terms).  Without implying that such calculi need
to be confluent, we remark that the system without the closed conditions
introduces non joinable critical pairs in the reduction rules resulting in
non-confluent systems. Notice that path computation in the GoI has only been
shown sound for nets that do not contain auxiliary doors.

The main results of this paper establish that the labels are adequate enough for
the representation of paths in proof-nets. This makes these calculi appealing for
intermediate representation of implementations of programing languages where
target compilation structures are linear logic proof-nets.

For the study of shared reductions in proof-nets, and in the calculus itself, a
few additions would be useful: it would be certainly interesting to track not
only cuts that correspond to \emph{Beta}-redexes but also exponential cuts
corresponding to substitutions.  For this, we should allow copy, erase and
substitution terms to bear labels. These additions could also be useful towards
obtaining a standardisation result for closed reduction calculi.

\clearpage
\appendix 
\begin{center}
   {\bf APPENDIX}
\end{center}

\newtheorem{theoremEx}{Theorem}
\renewcommand{\thetheoremEx}{A-\arabic{theoremEx}}
\newtheorem{lemmaEx}{Lemma}
\renewcommand{\thelemmaEx}{A-\arabic{lemmaEx}}

\section*{Proof of Theorem~\ref{th:main1}}
\begin{theoremEx}
  Let $W_G=\{w(\phi) \mid \phi \in \mbox{straight paths of } G\}$
  denote the set of weights of straight paths observable in a graph $G$ and let $T$ be a
  labelled term. If $T \to T'$ then $W_{G(T)} = W_{G(T')}$.
\end{theoremEx}

\begin{proof} By induction on $T$. The only interesting case is when
  the reduction takes place at the root position. We show the property
  by cases on the rule applied.

  \textit{Case \emph{Beta}}: We give the graphical representation of
  the left- and right-hand
  sides of the \emph{Beta}-rule below:\\
  \centerline{
    \includegraphics[width=0.9\textwidth]{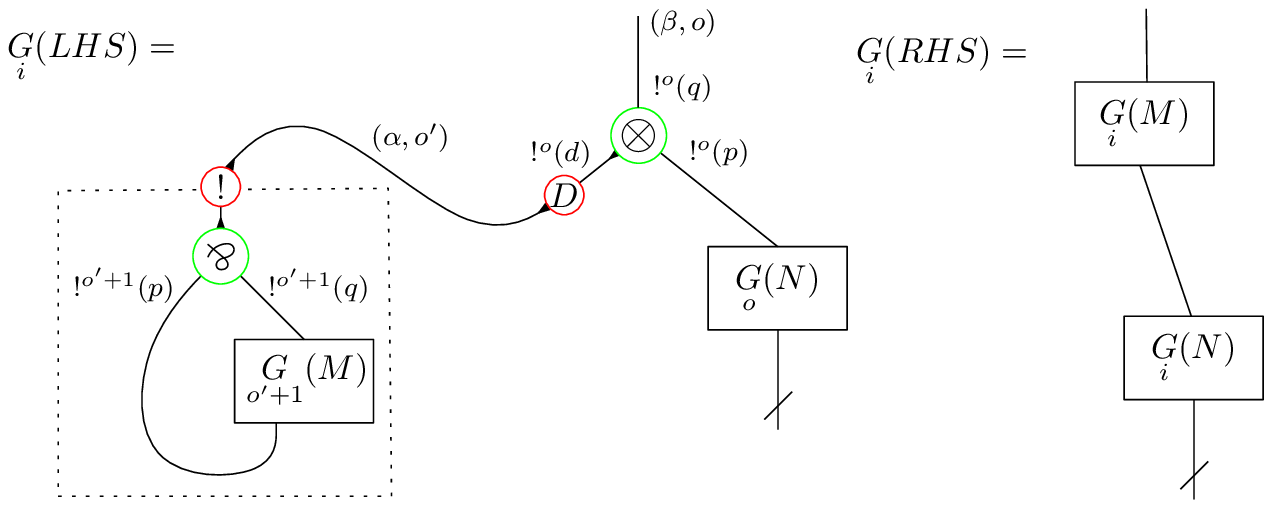}}
  Notice that there are no auxiliary doors since the function must be
  closed. We must show that weights of straight paths in the left hand
  side are found in the right hand side, that is, the weights of
  \begin{enumerate}
  \item $ \phi = \rho - (\beta,o) - !^o(q) - !^o(d) - (\alpha,o') -
    !^{o'+1}(q^*) - \mu $ where $\rho$ is a path ending at the root of
    the left hand side and $\mu$ is a path starting in $M$;
  \item $\psi = \mu' - !^{o'+1}(p) - (a^*, o') - !^{o}(d^*) -
    !^{o}(p^*) - \nu$ with $\mu'$ ending in the (translation of) free
    variable of $M$ and $\nu$ starting in $N$; and
  \item$\psi^r, \phi^r$
  \end{enumerate}
  are found in the right hand side of the figure.  Notice that in the
  right hand side, the translation of a substitution term does not
  place any label at the immediate root of the graph and there seems
  to be a kind of mismatch with the level numbers in which the
  subgraphs are called, however this is correct. Since we do not have
  labels on substitution terms, the translation will respect the $i's$
  later on, that is, we can apply the induction hypothesis.  We show
  the property by checking that the external label of $M$ fixes the
  level number yielding the weight we are after.
  \begin{itemize}
  \item We have $l= label ~ M = \beta\overline{\overrightarrow{\sf
        D}\alpha\overleftarrow{\sf!}}\cdot \beta'$, where $\beta'$ is
    some suffix. The weight of $l$ is given by
   $$\elstar[\beta\overline{\overrightarrow{\sf
       D}\alpha\overleftarrow{\sf!}}\beta' ] i = w((\beta,o) - !^o(q)
   - !^o(d) - (\alpha,o') - !^{o'+1}(q^*)) \cdot
   \elstar[\beta']{o'+1} $$ which completes the first case.
 \item We work in a similar fashion with the second case where we
   first translate $label ~ N=\underline{(\overrightarrow{\sf
       D}\alpha\overleftarrow{\sf!})^{r}}\gamma$.  Under the
   assumption that the wire connecting the free variable $x$ is at
   level $o+1$ we have
   $$\elstar[\underline{(\overrightarrow{\sf
       D}\alpha\overleftarrow{\sf!})^{r}}\gamma] i = w(!^{o'+1}(p) -
   (a^*, o') - !^{o}(d^*) - !^{o}(p^*)) \cdot \elstar[\gamma] o$$ The
   translation calls with a suitable $i$ big enough to cover the open
   scopes at the first label and returns the open scopes at the last
   label.  Now in $\psi$, reading $a$ in reverse means that the
   indicated scope number is the one for the first label and hence
   decreases.
 
   One argues in the same way for the reverse cases of $\phi$ and
   $\psi$.
 \end{itemize}

 \textit{Case \emph{Var}:} This is where two paths meet and get glued
 via an axiom-link.  The translation of the left- and right-hand sides
 do not tell us anything useful since we have just wires (identities)
 and thus the weights of paths remain the same.

 \textit{Case \emph{Lam:}} The translations of each side are:
 \\\centerline{
   \includegraphics[width=0.88\linewidth]{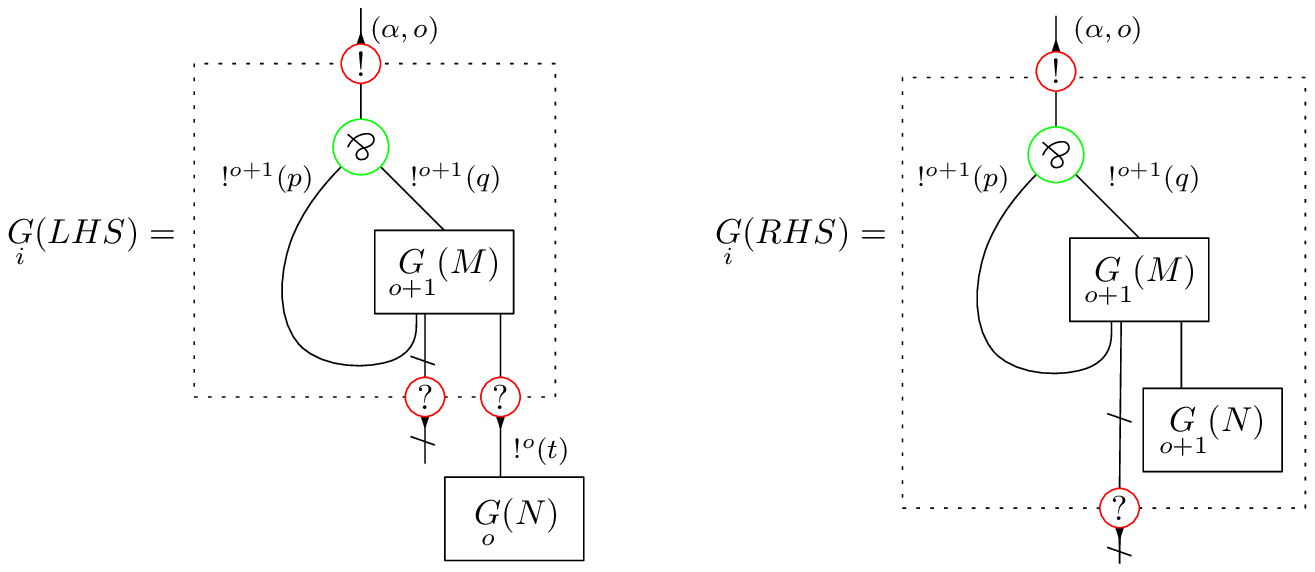} 
 } Recall that the external label of $N$ may give us a prefix of
 exponential markers followed by an underline or just an
 underline. The interesting point is that the right hand side suggests
 that the translation is called with $o+1$ for $N$ and looks like a
 source of a mismatch. But the external label on $N$ in the rhs must
 start with an exponential marker $?$ and this decreases the $o$ upon
 which $N$ is translated. Thus the weights of paths remain the same
 and this completes the case.

 \textit{Case \emph{Cpy1}:} We have $ (\delta_{x}^{y,
   z}.M)[N/x]\cfmer M[\overrightarrow{\sf R}\bullet
 N/y][\overrightarrow{\sf S}\bullet N/z]$ with the corresponding
 translations \\\centerline{
   \includegraphics[width=0.7\textwidth]{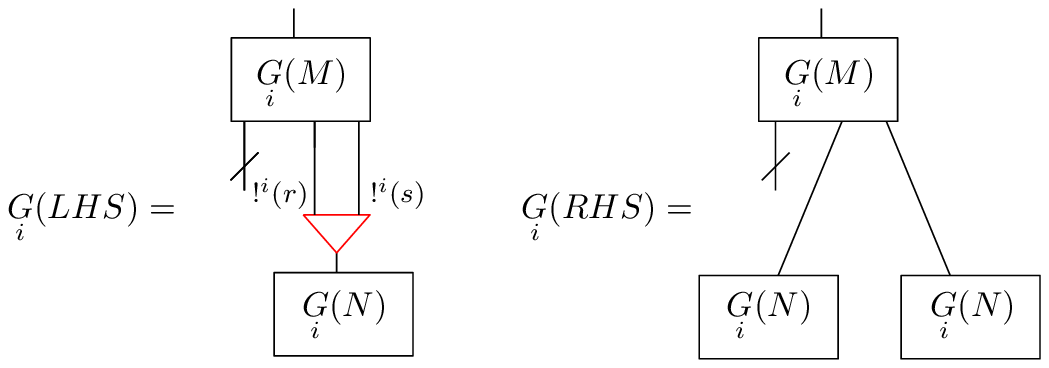}} We argue as before
 and there is nothing to say about the levels. The case is similar for
 erasing. 

 \textit{Case \emph{Ers1}:} The rule behaves as follows \\\centerline{
   \includegraphics[width=0.7\textwidth]{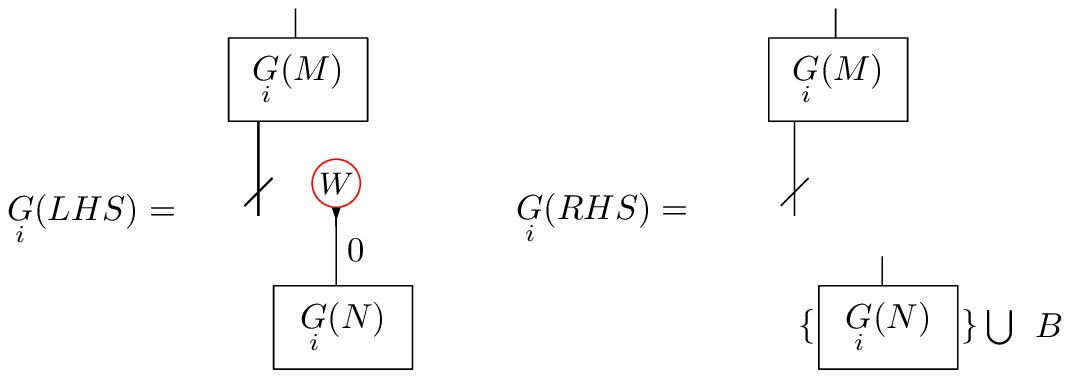}
 } There are killed paths in the left hand side and we have the same
 on the right.  Note that erased paths do not survive reduction but
 one may walk these with a strategy. This is the reason for keeping
 the set $B$, which is initially empty.

 The translation of rules App1, App2, Cpy2, Ers2 and Cmp correspond to
\end{proof}

\section*{Proof of Theorem~\ref{thm:cbn}}

\begin{lemmaEx}\label{lem:omega} If $T = M[N/x]$ then there is a decomposition
  $(label\ N) = \omega \underline{\alpha} \bright \sigma $ such that
  $\omega$ is a prefix built with exponential markers having the shape
  $\overrightarrow{E}_1 \dots \overrightarrow{E}_n $ where $n \ge 0$
  and $E$ is not a $D$ marker.
\end{lemmaEx}
\begin{proof}
  Since the term $N$ belongs to a substitution term, its external
  label must have been generated by a \emph{Beta}-rule generating the
  sub-label $\underline{\alpha} \bright \sigma$. By using the rules
  in $\sigma$, we can generate only an $\omega$ prefix. It cannot
  contain a dereliction marker since this can come  from the
  \emph{Var}-rule which stops the process;  that is, the root of $T$ is
  not a substitution term anymore.
\end{proof}
The consequence of the lemma is that derelictions are followed
directly by exponentials. Notice that this is different form the
previous system.

\begin{theoremEx} 
\begin{enumerate}
\item If $T \to_{lca} T'$ then $W_{G(T)} = W_{G (T')}$.
\item Suppose $T$ is a term obtained by erasing the labels and
  $\to_{ca}$ is the system generated by the rules for $\lambda_{lca}$
  by removing the labels.  If $T \to_{ca} T'$ then $G(T) \Rightarrow
  G(T')$ using closed cut elimination, where $G$ is the
  call-by-name translation.
\end{enumerate}
\end{theoremEx}

\begin{proof} We proceed by cases and argue about both properties.

  \textit{Case \emph{Beta}:} $\thebetalca$ where $\FV(N)=\emptyset$. The
  situation is the following:
  \\\centerline{
    \includegraphics[width=0.8\linewidth]{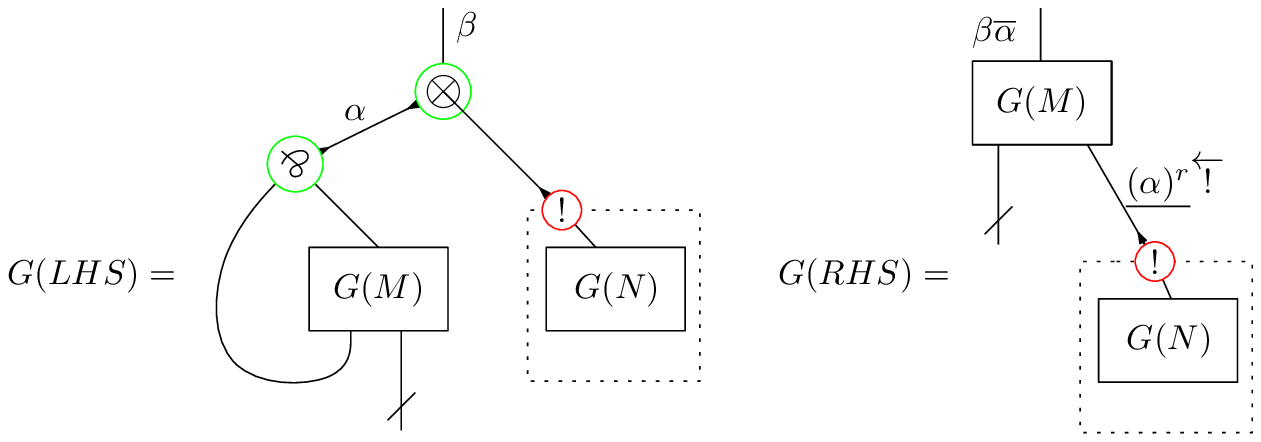}
  } Two paths are of interest: one entering from the root,
  moving along the cut and ending at the root of $G(M)$ and one coming
  from the free variable, travelling along the cut and ending up at
  the root of $G(N)$. Both weights are preserved in the right hand side.
  The second point of our claim is satisfied since the closed cut
  elimination sequence corresponds to one multiplicative cut.

  \textit{Case \emph{App2}:} $(MN)^{\alpha}[P/x]\camer
  (MN[\overrightarrow{\sf?}\bullet P/x])^{\alpha} $ where $ x\in
  \FV(N)$ and the translation of both sides become
  \\\centerline{ 
    \includegraphics[width=0.8\linewidth]{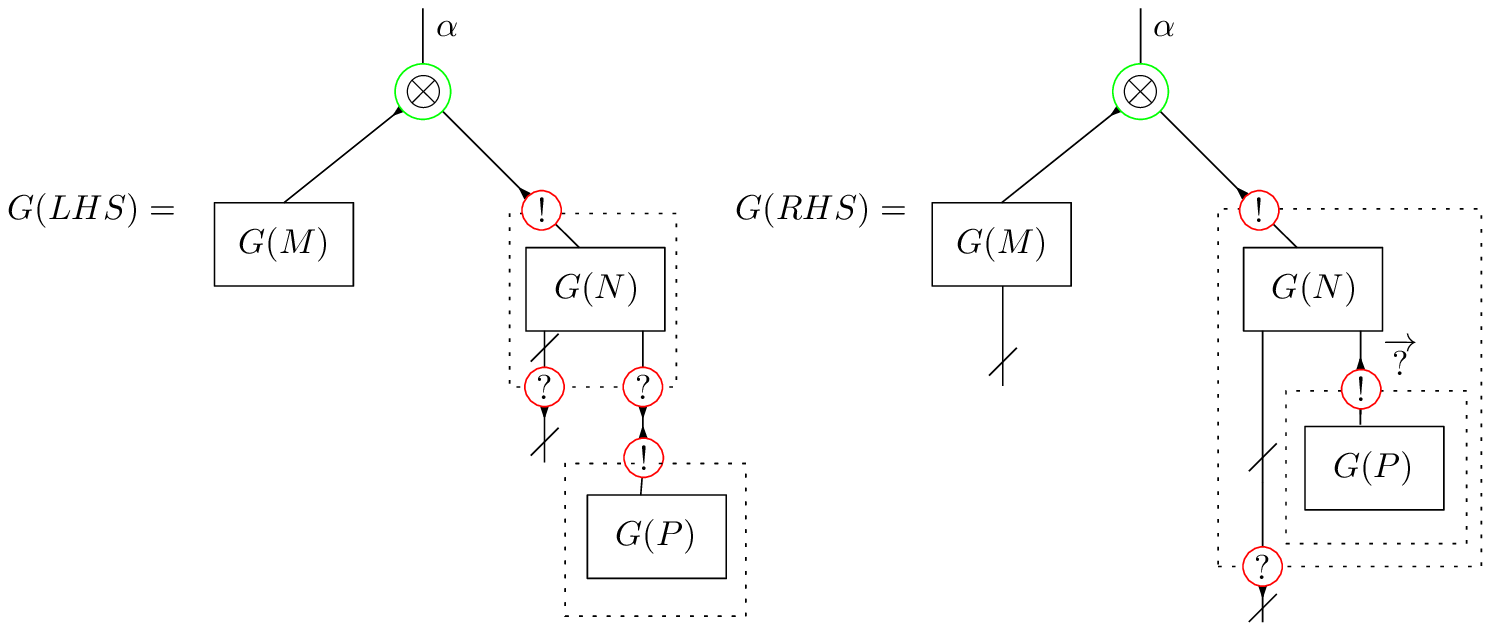} 
  }
  For the first point in our claim, weights are preserved by recording
  the auxiliary marker and for the second point, the graph rewrite
  corresponds to a closed commutative cut. Notice that there are no
  closedness conditions on the $\sigma$-rules, but since \emph{Beta}
  is the only rule that can generate a substitution, it must be a
  closed one.

  \textit{Case \emph{Var}:}
  $x^{\alpha}[N/x] \camer \alpha\overrightarrow{\sf D}\bullet N$ and
  the situation becomes
  \\\centerline{
    \includegraphics[width=0.8\linewidth]{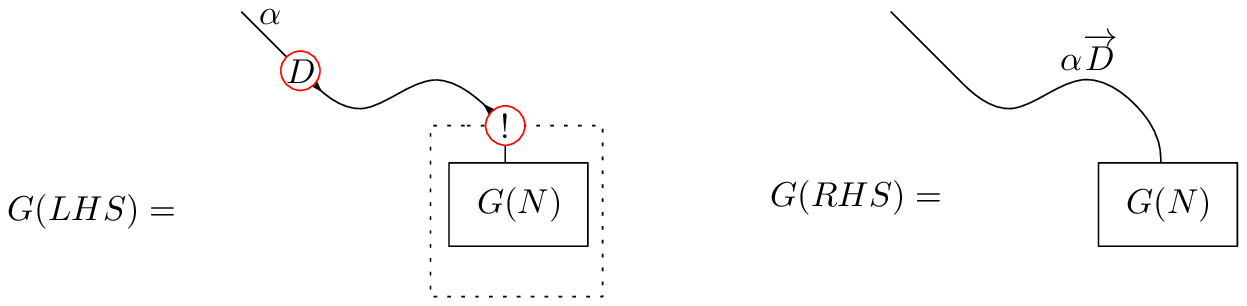}
  } There seems to be a mismatch with the graph rewriting
  rule and paths that we record since we removed the box in the right
  hand side.  However, by Lemma~\ref{lem:omega}, the external label of
  the argument must have the shape $(label\ N) = \omega
  \underline{\alpha} \bright \sigma $ such that $\omega$ is built with
  exponential markers and the trailing exponential box marker restores
  our level information. Regarding our second point, this simply
  corresponds to a closed dereliction cut.

  \textit{Case \emph{Cpy1, Ers1}:} With respect to the paths, both
  cases are similar as before. Regarding the second point of the
  claim, the translations correspond to a closed contraction and
  weakening cut respectively.

  For the remaining cases, the left and right hand side translations
  are identical.
\end{proof}
\end{document}